\documentclass[12pt,preprint]{aastex}

\slugcomment{ApJ Letters, {\bf 626}, L37, 2005}
\shorttitle{Spectrum of MHD Turbulence}
\shortauthors{Boldyrev}
\begin{document}
\input psfig.sty
\title{On the Spectrum of Magnetohydrodynamic Turbulence}
\author{Stanislav Boldyrev}
\affil{Department of Astronomy and Astrophysics, University of Chicago, 5640 South Ellis Avenue, Chicago, IL 60637}
\email{boldyrev@uchicago.edu}

\input psfig.sty

\begin{abstract}
We propose a phenomenological model for    
incompressible magnetohydrodynamic turbulence. We argue that 
nonlinear-wave interaction 
weakens as the energy cascade proceeds to small scales, 
however, the anisotropy of fluctuations along 
the large-scale magnetic field increases, which makes turbulence 
strong at all scales. 
To explain the weakening 
of the interaction, we propose that small-scale fluctuations of the 
velocity and magnetic fields become increasingly dynamically aligned  
as their scale decreases, so that 
turbulent ``eddies''  become locally anisotropic in the 
plane perpendicular  
to the large-scale magnetic field. In the limit of 
weak anisotropy, that is, weak large-scale magnetic field, our model reproduces 
the Goldreich-Sridhar spectrum, while the limit of strong anisotropy, that is, 
strong large-scale magnetic field,  corresponds 
to the Iroshnikov-Kraichnan 
scaling of the spectrum. This is in good agreement with recent  
numerical results. 
\end{abstract}
\section{Introduction.}
\label{introduction} 
Magnetohydrodynamic (MHD) turbulence is a state of a randomly stirred 
conducting fluid in the limit of very small fluid viscosity and resistivity;  
it plays an essential role in a variety of astrophysical systems, 
from planets and stars, to interstellar and 
intergalactic media~\citep[see, e.g.,][]{biskamp}. Despite more 
than 35 years of analytical, numerical, and observational investigations, 
the spectrum of MHD turbulence remains a subject of controversy. The standard 
results and recent developments in the theory of MHD turbulence can be found in 
many excellent texts \citep[see, e.g.,][]{iroshnikov,kraichnan,goldreich,goldreich2,bhattacharjee,cho,biskamp3,maron,vishniac,galtier1,galtier2,biskamp-muller,biskamp,kulsrud}. 
In the present Letter we first analyze the approaches of~\citet{iroshnikov} and  \citet{kraichnan} 
and \citet{goldreich} and point out the discrepancies of 
these theories with numerical results. 
Then, we propose a new model for the 
turbulent MHD cascade, which is free of these discrepancies, 
and which is in good agreement with recent high-resolution numerical findings   
of~\citet{maron}, \citet{biskamp-muller}, 
and~\citet{padoan}. 
Our results are summarized in Conclusions.\\

{\em The Iroshnikov-Kraichnan energy cascade.} 
The MHD equations describing evolution of the fluid velocity 
field, ${\bf v}({\bf r},t)$, and fluctuations of the  
magnetic field, ${\bf b}({\bf r},t)$, 
have an especially simple form when expressed in the Els\"asser variables, 
${\bf z}={\bf v}-{\bf b}$, and ${\bf w}={\bf v}+{\bf b}$:
\begin{eqnarray}
\partial_t {\bf z}+ ({\bf V}_A\cdot \nabla){\bf z}+({\bf w}\cdot \nabla){\bf z}=-\nabla P, 
\label{mhd1} \\
\partial_t {\bf w}-({\bf V}_A\cdot \nabla){\bf w}+({\bf z}\cdot \nabla){\bf w}=-\nabla P, 
\label{mhd2}
\end{eqnarray}
where the pressure $P$ is 
determined from the incompressibility 
condition, $\nabla \cdot {\bf z}=0$ or $\nabla \cdot {\bf w}=0$. In  
equations~(\ref{mhd1}) and (\ref{mhd2}), 
${\bf V}_{A}={\bf B}_{0}/\sqrt{4\pi \rho}$ is the Alfv\'en velocity, 
 $\rho$ is the fluid density, ${\bf B}_0$ is the large-scale 
external magnetic field, and we have neglected small viscosity and resistivity.   
Note that for ${\bf w}=0$, any function ${\bf z}={\bf f}({\bf r}-{\bf V}_A t)$ is a 
solution of the system~(eqs. \ref{mhd1},\ref{mhd2}); analogously, for ${\bf z}=0$, 
the solution is ${\bf w}={\bf g}({\bf r}+{\bf V}_A t)$, where the function~${\bf g}$ 
is arbitrary.  

Iroshnikov and Kraichnan used this fact to 
propose that the interacting Alfv\'en-wave packets (or ``eddies'') 
are those propagating in  
{\em opposite} directions along the large-scale 
magnetic-field lines~\citep{iroshnikov,kraichnan}. 
From that, they deduced that the energy transfer time from wave packets  
of size~$\lambda$ to smaller ones is increased  compared with the simple 
dimensional estimate $\tau(\lambda)\sim 
\lambda/\delta v_{\lambda}$. (We denote by $\delta v_{\lambda}$ and $\delta b_{\lambda}$ the velocity 
and magnetic-field fluctuations in the ``eddy.'' In the Alfv\'en wave, 
$\delta v_{\lambda}\sim \delta b_{\lambda}$.) Indeed, consider two wave packets of size~$\lambda$ 
propagating in the opposite directions along a magnetic-field line with the Alfv\'en velocities. 
Assuming that the ``eddies''   
are decorrelated at the field-parallel scale~$\lambda$, one can estimate from equations~(\ref{mhd1}) 
and (\ref{mhd2}) that during one collision the 
distortion of the ``eddy'' is $\Delta \delta v_{\lambda}\sim (\delta v_{\lambda}^2/\lambda)(\lambda/V_A)$. 
The distortions add up randomly, therefore, the ``eddy'' will be distorted relatively strongly after 
$N\sim (\delta v_{\lambda}/\Delta \delta v_{\lambda})^2\sim (V_A/\delta v_{\lambda})^2$ collisions. 
The energy transfer time is, therefore, 
\begin{eqnarray}
\tau_{IK}(\lambda)\sim N\lambda/V_A\sim  \lambda/\delta v_{\lambda}(V_A/\delta 
v_{\lambda}). 
\label{tauik}
\end{eqnarray} 

It is important that in 
the Iroshnikov-Kraichnan interpretation,   
an ``eddy'' has to experience {\em many} uncorrelated 
interactions with oppositely moving ``eddies'' before its 
energy is transferred to a smaller scale. Moreover,  turbulence 
becomes progressively {\em weaker} as the energy cascade proceeds toward 
smaller scales. The requirement of constant energy flux over 
scales $\delta v^2_{\lambda}/\tau_{IK}(\lambda)={\sf const}$ immediately 
leads to the scaling of fluid fluctuations $\delta v_{\lambda}\propto 
\lambda^{1/4}$, which resulted in the energy spectrum (the  
Iroshnikov-Kraichnan spectrum), 
\begin{eqnarray}  
E_{IK}(k)= |\delta v_k|^2k^2\propto k^{-3/2}.
\label{eik}
\end{eqnarray} 
Iroshnikov and Kraichnan did not consider anisotropy of the 
spectrum, so~$E_{IK}(k)$ was assumed to be 
three-dimensional and isotropic.\\

{\em Anisotropy: The Goldreich-Sridhar energy cascade.} 
Over the years, isotropy of the MHD spectrum 
in a strong external magnetic 
field seemed to contradict analytical and numerical findings, \citep[see, e. g.,][]{biskamp}.   
A theory of anisotropic MHD turbulence was proposed by~\citet{goldreich}. 
They argued that  ``eddies'' 
are strongly anisotropic; they are elongated along the large-scale 
magnetic field lines. As a consequence, the time of relatively 
strong ``eddy'' distortion (the energy transfer time) is on the order 
of a crossing time required 
for two oppositely moving ``eddies'' to pass each other. 

Suppose that the ``eddy'' has a transverse  
(to the large-scale field) size~$\lambda$. Then, its field-parallel size~$l$
can be found from the so-called critical-balance condition, proposed by~\citet{goldreich}.  
This condition has two 
explanations that are equivalent  
in the Goldreich-Sridhar picture. First, the critical balance 
can be understood as a formal balance 
of the second and third terms in equations~(\ref{mhd1}) and (\ref{mhd2}): 
$V_A/l\sim \delta b_{\lambda}/\lambda$.  Second, it follows from the geometric 
distortion of magnetic-field lines in the turbulent ``eddy.'' Indeed, 
the ``eddy'' displaces the lines  
in their perpendicular direction by a distance~$\xi\sim \delta b_{\lambda}l/V_A $, and this 
displacement equals the perpendicular ``eddy'' size~$\lambda$. 

The critical balance condition is the same for 
all scales, so, contrary to the Iroshnikov-Kraichnan picture, the turbulence strength 
does not change with the scale. 
The energy transfer time predicted in the Goldreich-Sridhar theory is 
\begin{eqnarray}
\tau_{GS}(\lambda)\sim l/V_A\sim \lambda/\delta v_{\lambda}.   
\label{taugs}
\end{eqnarray}
Assuming that the energy cascade is independent of the scale, 
$\delta v^2_{\lambda}/\tau_{GS}(\lambda)={\sf const}$, one obtains the 
scaling of fluid fluctuations,   
$\delta v_{\lambda}\propto \lambda^{1/3}$. 
The corresponding energy spectrum is
 \begin{eqnarray}
E_{GS}(k_{\perp})=|\delta v_{k_{\perp}}|^2k_{\perp}\propto k_{\perp}^{-5/3}. 
\end{eqnarray}
This spectrum coincides with the Kolmogorov spectrum of incompressible 
non-magnetized fluid turbulence, as it should since the energy transfer time 
coincides with the ``eddy'' turn-over time, 
$\tau(\lambda)\sim \lambda/\delta v_{\lambda}$, in both approaches. 
The anisotropy of  fluctuations is described by the 
condition that follows from the critical balance, $l\propto \lambda^{2/3}$. 
One can 
therefore write that the fluctuations 
scale with the field-parallel size of the ``eddy'' 
as~$\delta v_l\propto l^{1/2}$. 

The Goldreich-Sridhar picture, however, does not fully agree with  
numerical simulations. As was recently discovered     
by~\citet{biskamp-muller}, the anisotropic spectrum 
depends on the strength of the external magnetic field. Denote 
$\gamma=B^2_0/ \langle\rho \delta v^2_L\rangle$, 
where $\delta v_L$ is the velocity field at the outer scale of 
turbulence,~$L$. It was   
found that the {\em field-perpendicular} scaling of fluctuations changed from 
the Goldreich-Sridhar form to the Iroshnikov-Kraichnan form as the 
field increased from~$\gamma \ll 1$ to~$\gamma\gg 1$. 
A similar result for $\gamma\gg 1$ was obtained 
earlier by~\citet{maron}.  
These intriguing numerical findings motivated our interest in the 
problem. 
In the next section, we propose a phenomenological 
model of MHD turbulence, which agrees well with available numerical results, 
for any strength of the external field. In the limiting case of a weak external field, 
our model reproduces the anisotropic spectrum of Goldreich \& Sridhar.  
In the other limiting 
case of a very strong external field, anisotropy of the 
spectrum is {\em stronger};  
however, the {\em field-perpendicular} spectrum formally coincides 
with the spectrum predicted in the Iroshnikov-Kraichnan model.

\section{A model for MHD turbulence.}
\label{model}
To begin with, we make a certain assumption about reduction 
of the nonlinear interaction, 
which is not present in either the Iroshnikov-Kraichnan or Goldreich-Sridhar picture. We postpone the  
justification of this assumption until the end of this section, 
when we obtain the corresponding solution for the turbulent spectra and compare it  
with numerical simulations. \\ 

{\em Reduction of nonlinear interaction.} 
Let us assume that the nonlinear interaction of the 
counter-propagating 
fluctuations is reduced by a factor $(\delta v_{\lambda}/V_A)^{\alpha}$, 
where $\alpha$ is some undetermined exponent, $0\leq\alpha \leq 1$. 
In other words, we assume 
that the nonlinearity in equations~(\ref{mhd1}) and (\ref{mhd2}) is ``depleted,'' so that 
the interaction terms are of order 
\begin{eqnarray}
({\bf w}\cdot \nabla){\bf z}
\sim ({\bf z}\cdot \nabla){\bf w}
\sim (\delta v^2_{\lambda}/\lambda)(\delta v_{\lambda}/V_A)^{\alpha}. 
\label{reduction}
\end{eqnarray}
Thus, the fluid 
fluctuations at the transverse distance $\lambda$ become decorrelated on  
the time-scale 
\begin{eqnarray}
\tau_N(\lambda)\sim (\lambda/\delta v_{\lambda})
\left(V_A/\delta v_{\lambda}\right)^{\alpha}.
\label{taun}
\end{eqnarray}
Their decorrelation length along the magnetic field line can be 
found from the causality principle. For $\delta v_{\lambda}< V_A$,  
the perturbation cannot propagate 
along the field line faster than $V_A$; therefore, the correlation length 
along the field line is on the order of $l\sim V_A\tau_{N}(\lambda)$. 
This condition is analogous to the critical balance condition of the 
previous section in that it satisfies the same balance between 
the linear and nonlinear 
interaction terms in equations~(\ref{mhd1}) and (\ref{mhd2}). However, its 
geometric meaning is different, and will be discussed below.
  
We see that the interaction strength~(eq. \ref{reduction}) decreases for smaller scales; however, 
the field-parallel ``eddy'' size~$l$   
is adjusted in such a way 
that the energy transfer to the smaller scales always takes 
one ``eddy'' crossing time. Note that contrary 
to the Iroshnikov-Kraichnan formalism, and similar to the Goldreich-Sridhar approach, 
in our picture turbulence is strong  and fluctuations are highly 
anisotropic. Now, we require that the energy cascade be constant over 
scales $\delta v^2_{\lambda}/\tau_N(\lambda)={\sf const}$. We 
derive $\delta v_{\lambda}\propto \lambda^{1/(3+\alpha)}$, 
and the anisotropy condition reads $l\propto \lambda^{2/(3+\alpha)}$. The 
corresponding energy spectrum is given by
\begin{eqnarray}
E(k_{\perp})= |\delta v_{k_{\perp}}|^2k_{\perp}\propto 
k^{-(5+\alpha)/(3+\alpha)}_{\perp}.
\end{eqnarray}
One can formally define the corresponding longitudinal  spectrum of fluctuations, 
from the condition 
$E(k_{\perp})d k_{\perp}=E(k_{||})d k_{||}$, with 
$k_{\perp}\propto k_{||}^{(3+\alpha)/2}$. The answer is 
$E(k_{||})\propto k^{-2}_{||}$; note that it is independent 
of~$\alpha$. Therefore, the  scaling of fluid 
fluctuations with respect to the field-parallel distance~$l$  is always~$\delta v_{l}\propto l^{1/2}$.\\

{\em Comparison with numerical simulations.} 
Obviously, our result with $\alpha=0$ corresponds to the 
Goldreich-Sridhar scaling, while $\alpha =1$ produces 
the Iroshnikov-Kraichnan scaling.  
Simulations by~\citet{biskamp-muller}  
for a range of large-scale external magnetic fields have shown that the scaling 
of the second-order structure function with respect to 
the field-perpendicular scale~$\lambda$  
changes from the Goldreich-Sridhar value 
in the case of a weak external field, $\gamma\ll 1$, 
to the Iroshnikov-Kraichnan value in the case of 
a strong field, $\gamma \gg 1$. At the same time, the scaling 
of the second-order structure 
function with respect to the field-parallel distance~$l$ 
does not change much and stays close 
to $\langle (\delta v_{l})^2\rangle\propto l^{0.9}$. This result  
is consistent 
with our prediction, $(\delta v_{l})^2\propto l$. The same scalings 
for the case of a 
strong external magnetic field, were earlier observed in 
simulations by~\citet{maron}. In these simulations, 
it was also found that the scaling of the energy-cascade 
time was $\tau_E\propto \lambda^{1/2}$. 
This contradicts the Goldreich-Sridhar picture \citep[see the discussion 
in \S\, 6.1.3 of][]{maron}, 
but coincides with our formula $\tau_N(\lambda)\propto \lambda^{(1+\alpha)/(3+\alpha)}$, 
for~$\alpha=1$. 

It is important to note that when the perpendicular Iroshnikov-Kraichnan 
scaling  
is formally reproduced in our model, the turbulent fluctuations are strong 
and anisotropic, with~$l\propto \lambda^{1/2}$. 
Moreover, the turbulence possessing 
the Iroshnikov-Kraichnan spectrum is more anisotropic 
than the turbulence in the Goldreich-Sridhar model, which is 
not surprising, since, as just discussed, the Iroshnikov-Kraichnan 
spectrum corresponds to a much stronger 
external magnetic field.

{\em Scale-dependent dynamic alignment.} 
We next explain a  
possible physical origin for  
the proposed reduction of the non-linear interaction~(\ref{reduction}). 
A hint toward the explanation can be obtained from  geometric 
considerations. The displacement of magnetic field lines in 
the direction  perpendicular to the large-scale magnetic field, 
produced by the 
wave packet with $l\propto \lambda^{2/(3+\alpha)}$ and $\delta b_{\lambda}\propto 
\lambda^{1/(3+\alpha)}$, is given by $\xi\propto \delta b_{\lambda} l\propto 
\lambda^{3/(3+\alpha)}$. We thus obtain that 
the transverse displacement of magnetic field lines 
in the shear-wave packet is on the 
order of~$\xi$ at distances~$\lambda \ll \xi$. 
Therefore, this packet should be highly anisotropic in the 
plane {\em perpendicular} to the large-scale magnetic field.  In this plane, 
it is elongated in the direction of the field fluctuations.  
This may be consistent with the numerically supported 
picture that dissipative 
structures in MHD turbulence are micro-current sheets~\citep{biskamp2,biskamp,biskamp3,maron}. 
For comparison,   
the Goldreich-Sridhar scaling, $\alpha=0$, would correspond to 
filament-like dissipative 
structures. In Figure~\ref{eddies} we sketch the shapes of the ``eddies'' in the Goldreich-Sridhar  
picture and in our picture. In our model, the anisotropy of the ``eddies'' 
is weakest for $\alpha=0$, and therefore it is natural 
to identify this limit with zero 
large-scale magnetic field. In the other limit, $\alpha=1$, the anisotropy of 
the ``eddies'' is strongest, therefore, this case should be 
identified with a strong magnetic field,~$\gamma \gg 1$.~\footnote{As is 
shown in~\citep{grappin} with the aid of the EDQNM closure, strong large-scale 
magnetic field leads to the Alfv\'enic decorrelation time in aligned 
turbulence. In our model, this means that $\alpha$ cannot exceed~$1$.} 

Let us now explain how the envisioned  
anisotropy of fluctuations in the field-perpendicular plane 
reduces the strength 
of the nonlinear-wave interaction. For this purpose, consider the 
third terms in the MHD equations (\ref{mhd1},\ref{mhd2}), 
say, $({\bf w}\cdot \nabla){\bf z}$. Since both  
${\bf w}$ and ${\bf z}$ are divergence-free and perpendicular to the 
large-scale magnetic field, this term is proportional 
to the angle~$\theta_{\lambda}$ between the directions of~${\bf w}_{\lambda}$ 
and~${\bf z}_{\lambda}$ 
(if this angle is small). But we just established that this angle 
is $\theta_{\lambda}\sim \lambda/\xi\propto \lambda^{\alpha/(3+\alpha)}$. 
Quite remarkably, we have reproduced the reduction 
factor $(\delta v_{\lambda}/V_A)^{\alpha}$ 
in equation~(\ref{reduction}). This demonstrates that our 
initial assumption is self-consistent.  Moreover, a detailed analysis of 
equations~(\ref{mhd1}) and (\ref{mhd2}) suggests that the 
alignment $\theta_{\lambda}\propto \delta v_{\lambda}^{\alpha}$ is indeed  
consistent with the MHD dynamics, when $\theta_{\lambda}$ is small. In this 
limit we can obtain the following approximate evolution equations, 
$\partial_t \delta v_{\lambda}\sim \delta v_{\lambda}^2\theta_{\lambda}/\lambda$ and 
$\partial_t \theta_{\lambda}\sim \delta v_{\lambda}\theta^2_{\lambda}/\lambda$. 
The alignment exponent 
${\alpha}$ would be determined by numeric coefficients in these equations, 
which cannot be obtained from the dimensional analysis.

As an important analogy, we 
note that {\em decaying} 
MHD turbulence approaches the so-called Alfv\'enic state, where 
either~${\bf w}$ or~${\bf z}$ is zero depending on the initial 
conditions \citep[see e.g.,][]{grappin,grappin2}.    
Based on our analysis, we propose that driven turbulence behaves in 
a similar manner, although~${\bf b}$ does not become exactly 
equal to~$\pm {\bf v}$. Rather, magnetic 
fluctuations $\delta {\bf b}_{\lambda}$ 
tend to align 
their direction, but not their magnitude, with that of the velocity 
fluctuations, $\pm \delta {\bf v}_{\lambda}$. As a result, the  
fluctuations $\delta {\bf w}_{\lambda}$ and $\delta {\bf z}_{\lambda}$ are of the 
same order, and the directions of  $\delta {\bf w}_{\lambda}$ and $\pm \delta {\bf z}_{\lambda}$ 
are aligned within the angle 
$\theta_{\lambda}\sim \lambda/\xi\propto \lambda^{\alpha/(3+\alpha)}$.  
The degree of the alignment increases progressively as the scale decreases.  
Such scale-dependent dynamic alignment (and the associated depletion of nonlinearity) 
can in principle 
be checked numerically, although the numerical analysis may be 
complicated by the rather slow dependence of the alignment angle~$\theta_{\lambda}$ on 
the scale. There is, however, a numerical indication that MHD turbulence indeed 
has a tendency to create correlated regions of polarized 
fluctuations \citep{maron}.

{\em On the non-universality of the turbulent spectrum.}
Our analysis points to an interesting possibility that 
 MHD turbulence is non-universal in that it depends on the large-scale magnetic field. 
We may further speculate that, 
in principle, other large-scale 
conditions may affect the scaling properties of turbulence. 
For example, the dynamic alignment may be sensitive to the level of 
cross-helicity fluctuations; an analogous result is known for the case 
of {\em decaying} turbulence, \citep[see, e.g.,][]{grappin,grappin2}. An alternative 
possibility is that the spectrum is {\em universal} and has the Iroshnikov-Kraichnan 
scaling but, for $\gamma \ll 1$, the dynamic alignment with $\alpha=1$ is established 
rather slowly as the scale decreases, and the resolution of numerical 
simulations is not large enough to reach the universal regime.

\section{Conclusions.}
\label{conclusions}
To conclude, we summarize our main results: 

1. Our consideration is motivated by the recent numerical observations 
\citep{maron,biskamp-muller}  
that incompressible MHD turbulence is not completely 
described  by either the Iroshnikov-Kraichnan or the Goldreich-Sridhar model. 
The scaling of velocity 
fluctuations was found in these papers to depend on the 
strength of the large-scale external 
magnetic field: the Iroshnikov-Kraichnan scaling 
appeared in the limit of strong 
external magnetic field, while the Goldreich-Sridhar scaling appeared in 
the limit of weak field.

2. To explain these numerical findings, we propose that turbulent 
fluctuations 
become increasingly dynamically aligned as the energy cascade proceeds 
to smaller scales. Velocity fluctuations 
$\delta {\bf v}_{\lambda}$ tend to align their direction with that of 
magnetic field fluctuations, $\pm\delta {\bf b}_{\lambda}$, 
and the smaller the scale~$\lambda$, the stronger the alignment. 
The dynamic alignment leads to reduction of the nonlinear-wave interaction 
(so-called depletion of nonlinearity). 

\begin{figure} [tbp]
\centerline{\psfig{file=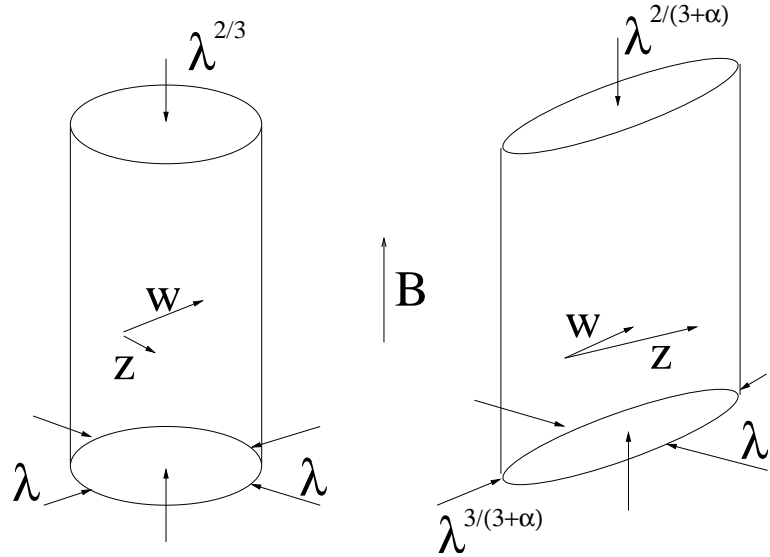,width=4.in,angle=0}}
\vskip5mm
\caption{A sketch of the shapes of turbulent fluctuations, ``eddies,'' in the Goldreich-Sridhar 
picture (left), and in our picture (right). The large-scale magnetic field is denoted by~${\bf B}$.
As the turbulent cascade proceeds toward the smallest, 
dissipative scales,~$\lambda \to 0$, the Goldreich-Sridhar ``eddy'' assumes the shape of a filament, while the 
``eddy'' predicted in our model turns into a current sheet.}
\label{eddies}
\end{figure}

3. As a result of point 2,  fluctuating ``eddies''  are three-dimensionally  
anisotropic. The ``eddies,'' 
whose smallest scale is $\lambda$, have the scale $\xi\propto\lambda^{3/(3+\alpha)}$ 
in the direction of the shear (the direction of the magnetic field line 
distortion), and the scale $l\propto \lambda^{2/(3+\alpha)}$ in the 
direction of 
the large-scale magnetic field, as is sketched in Figure~\ref{eddies}. The scaling and 
anisotropy of  fluctuations are described by a single 
parameter~$\alpha$, which depends on the strength of the external magnetic field, and 
which is determined in our work from comparison with numerical simulations 
by~\citet{maron} and \citet{biskamp-muller}.

4. The energy distribution is given by 
$$E(k_{\perp})\propto k_{\perp}^{-(5+\alpha)/(3+\alpha)}.$$ 
This coincides with the Goldreich-Sridhar spectrum in the limit of 
weak anisotropy~($\alpha=0$), and with 
the Iroshnikov-Kraichnan spectrum in the limit of strong 
anisotropy~($\alpha=1$).  As the external magnetic field increases 
from~$\gamma \ll 1$ to~$\gamma\gg 1$, 
the anisotropy of turbulent fluctuations increases, and 
the parameter~$\alpha$ 
changes from~$\alpha=0$ to~$\alpha=1$. According to point 3, 
the corresponding scalings 
of the fluctuations with respect to their field-perpendicular 
and field-parallel dimensions 
are $\delta v_{\lambda}\propto \lambda^{1/(3+\alpha)}$, $\delta v_{\xi}\propto \xi^{1/3}$,  
and $\delta v_{l}\propto l^{1/2}$. 

5. The smallest-scale ``eddies'' in our turbulent 
cascade ($\lambda \to 0$ for $\alpha \neq 0$) have 
a sheet like morphology,  in agreement with micro current-sheet 
dissipative 
structures, numerically observed in MHD turbulence~\citep{biskamp3}. 
In the 
case of zero external magnetic field, $\alpha=0$, the dissipative 
structures are filaments, which also agrees with numerical 
simulations~\citep{padoan}.

6.  Previous attempts to explain the numerically observed 
Iroshnikov-Kraichnan scaling (and the associated cascade-time increase~(\ref{tauik})),   
in anisotropic sub-Alfv\'enic MHD turbulence ($\gamma \gg 1$) essentially invoked 
intermittency effects, \citep[see, e.g., \S\, 6.6.4 in][]{maron,biskamp-muller}.  
The intermittency effects are essential for explaining higher order 
statistics of MHD turbulence, that is, scaling of  
higher order structure functions of~${\bf w}$ and~${\bf z}$. However, they  
usually provide only small corrections to the scaling of the second-order structure functions, and 
of the energy spectra. Such effects are not addressed in the present Letter; 
our derivation is based on the idea of scale-dependent dynamic 
alignment, which does not require intermittency.

\acknowledgments
I am grateful to Fausto Cattaneo and Samuel Vainshtein for many discussions of MHD 
turbulence, to Jason Maron, the referee of 
this Letter, for important comments on the results of~\citet{maron}, 
to Wolf-Christian M\"uller for discussion of the results 
of~\citet{biskamp-muller}, and to Dmitri Uzdensky for helpful comments on the physics and 
the style of the paper. 
This work was supported by the NSF Center for Magnetic
Self-Organization in Laboratory and Astrophysical Plasmas
at the University of Chicago. The hospitality of the Aspen Center for Physics, 
where part of this work was done, is gratefully acknowledged.

\newpage

\begin{thebibliography}{99}

\bibitem[Biskamp(2003)]{biskamp} Biskamp, D. {\em Magnetohydrodynamic Turbulence.} 
(Cambridge University Press, Cambridge, 2003).
\bibitem[Biskamp(1993)]{biskamp2} Biskamp, D. {\em Nonlinear Magnetohydrodynamics.} (Cambridge Universtiy Press, Cambridge, 1993). 
\bibitem[Biskamp \& M\"uller(2000)]{biskamp3} Biskamp, D. \& M\"uller, W.-C., Phys. Plasmas {\bf 7} (2000) 4889.
\bibitem[Cho \& Vishniac(2000)]{cho} Cho, J. \& Vishniac, E., ApJ., {\bf 539} (2000) 273.
\bibitem[Cho, Lazarian \& Vishniac(2002)]{vishniac} Cho, J., Lazarian, A., \& Vishniac, E., ApJ {\bf 564} (2002) 291.
\bibitem[Galtier, et al.(2000)]{galtier1} Galtier, S., Nazarenko, S. V., Newell, A. C., \& Pouquet, A., 
J.~Plasma Physics {\bf 63} (2000) 447.
\bibitem[Galtier, et al.(2002)]{galtier2} Galtier, S., Nazarenko, S. V., Newell, A. C., \& Pouquet, A., 
ApJ {\bf 564} (2002) L49.
\bibitem[Goldreich \& Sridhar(1995)]{goldreich} Goldreich, P. \& Sridhar, S., ApJ {\bf 438} (1995) 763.
\bibitem[Goldreich \& Sridhar(1997)]{goldreich2} Goldreich, P. \& Sridhar, S., ApJ {\bf 485} (1997) 680.
\bibitem[Grappin, et al(1982)]{grappin} Grappin, R., Frisch, U., L\'eorat, J., \& Pouquet, A.,  
Astron. Astrophys.~{\bf 105} (1982) 6.
\bibitem[Grappin, Pouquet, \& L\'eorat(1983)]{grappin2} Grappin, R., Pouquet, A., \& L\'eorat, J.,  
Astron. Astrophys.~{\bf 126} (1983) 51.
\bibitem[Iroshnikov(1963)]{iroshnikov} Iroshnikov, P. S.,  AZh, {\bf 40} (1963) 742; Sov. Astron, {\bf 7} (1964) 566.
\bibitem[Kraichnan(1965)]{kraichnan} Kraichnan, R. H.,  Phys. Fluids, {\bf 8} (1965) 1385.
\bibitem[Kulsrud(2005)]{kulsrud} Kulsrud, R. M., {\em Plasma Physics for Asrophysics.} 
(Princeton University Press, 2005).
\bibitem[Maron \& Goldreich(2001)]{maron} Maron, J., \& Goldreich, P., ApJ {\bf 554} (2001) 1175.
\bibitem[M\"uller, Biskamp \& Grappin(2003)]{biskamp-muller}  M\"uller, W.-C., Biskamp, D., \& Grappin, R., 
Phys. Rev. E~{\bf 67} (2003) 066302.
\bibitem[Ng \& Bhattacharjee(1996)]{bhattacharjee} Ng, C. S., \& Bhattacharjee, A., ApJ {\bf 465} (1996) 845. 
\bibitem[Padoan, et al.(2004)]{padoan} Padoan, P., Jimenez, R., Nordlund, \AA, \& Boldyrev, S.,  
Phys. Rev. Lett.~{\bf 92} (2004) 191102.


\end{thebibliography}
\end {document}